\documentclass[11pt]{article}
\usepackage{moriond,epsfig}
\usepackage{graphicx,changebar}
\usepackage{amsmath}
\def\leqsim{\mathbin{\;\raise1pt\hbox{$<$}\kern-8pt\lower3pt\hbox{$\sim$}\;}}
\def\geqsim{\mathbin{\;\raise1pt\hbox{$>$}\kern-8pt\lower3pt\hbox{$\sim$}\;}}

\def\MXN#1{\mbox{$ M_{\tilde{\chi}^0_#1}                                $}}

\def\XPM#1{\mbox{$ \tilde{\chi}^{\pm}_#1                                $}}
\def\XN#1{\mbox{$ \tilde{\chi}^0_#1                                     $}}

\def\p#1{\mbox{$ \mbox{\bf p}_1                                         $}}

\newcommand{\tanb}    {\mbox{$ \tan \beta                                  $}}
\newcommand{\mzero}   {\mbox{$ m_0                                         $}}
\newcommand{\mhf}     {\mbox{$ m_{3/2}                                     $}}
\newcommand{\sgnmu}   {\mbox{$ {\mathrm sign}(\mu)                         $}}

\newcommand{\snu}     {\mbox{$ \tilde\nu                                   $}}
\newcommand{\msnu}    {\mbox{$ M_{\tilde\nu}                               $}}

\newcommand{\stau}    {\mbox{$ \tilde{\tau}                                $}}

\newcommand{\ee}      {\mbox{$ {\, \mathrm e}^+ {\mathrm e}^-                 $}}

\newcommand{\GeVcc}   {\mbox{$ {\mathrm{GeV}}/c^2                          $}}

\newcommand{\TeVcc}   {\mbox{$ {\mathrm{TeV}}/c^2                          $}}

\newcommand{\etal}  {\mbox{\it et al.}}
\def\NPB#1#2#3{{\rm Nucl.~Phys.} {\bf{B#1}} (#2) #3}

\def\Journal#1#2#3#4{{#1} {\bf #2}, #3 (#4)}

\def\NPB{{\em Nucl. Phys.} B}

\def\be{\begin{equation}}
\def\ee{\end{equation}}
\def\bea{\begin{eqnarray}}
\def\eea{\end{eqnarray}}

\begin{document}
\title{Search for AMSB with the DELPHI data}

\author{T. Alderweireld}

\address{Facult\'e des Sciences, Universit\'e de  Mons Hainaut, Mons, Belgium \\ email: thomas.alderweireld@cern.ch}

\maketitle\abstracts{The data collected  by the DELPHI  experiment  up to the highest  LEP2 energies were used  to put  constraints on  the
  Anomaly Mediated SUSY
Breaking model   with   a flavour independent $m_0$ parameter.   The experimental searches covered  several possible signatures experimentally
 accessible at LEP,  
with either the lightest neutralino,  the sneutrino  or the stau  being  the  LSP.  They  included  the  search  for  nearly  mass degenerate 
 chargino  and  neutralino (always  present  in  AMSB), the
search for Standard Model like or invisible Higgs boson, the search for stable staus, and the search for cascade  decays  resulting in the LSP
(neutralino or sneutrino) and a low multiplicity final state containing neutrinos.}

\section{Introduction}\label{sec:intro}
Anomaly Mediated Supersymmetry Breaking (AMSB) \cite{branes,amsb} is an interesting solution to 
the flavour problem of mSUGRA. Rescaling anomalies in the supergravity Lagrangian always gives rise
to soft mass parameters in the observable sector. It follows that anomalies contribute to the SUSY
breaking in any case, whatever is the symmetry breaking mechanism. We'll refer to AMSB as the model 
in which all other components that mediate the SUSY breaking are suppressed, and the anomaly 
mediation is the dominant mechanism. 

AMSB is very predictive: all the low energy phenomenology can be derived by adding to the Standard
Model (SM) just two extra parameters and one sign. Unfortunately, the minimal AMSB model results in
tachyonic masses for sleptons at the electroweak scale. 
One way of getting rid of tachyons is to suppose additional, non anomaly, contributions to the SUSY
breaking which can generate a positive contribution ($m_0^2$) to the soft masses squared. It has to be emphasized that AMSB 
scenarios favor light Higgs ($h^0$), i.e. $m_{h^0} < 120 \ \GeVcc$, hence if the Higgs is not be found in the  runs 
at the Tevatron or, further on, at the LHC, the AMSB model itself will be completely 
ruled out.

\section{Phenomenology of AMSB}
\label{par:phenomenology}

If there is only one common squared mass term for all scalars, all masses and couplings can be 
derived in terms of just three parameters and one sign, namely, the mass of the gravitino, \mhf, 
the ratio of Higgs vacuum expectation values, \tanb , the common scalar mass parameter \mzero\ and the sign of the 
Higgs term, \sgnmu. In the model considered here, only the slepton mass spectrum and, to some extent,
 the Higgs depend on the assumptions of a common scalar term \mzero. All other features are characteristics of any AMSB 
scenario, whatever is the procedure used to cope with the tachyonic masses of the sleptons.
Since \mzero\ is a free parameter, according to its value there are three possible candidates for
the LSP: either the nearly mass degenerate \XN{1}/\XPM{1}, the \snu\ or the \stau. Scenarios with any of the above as LSP were explored using the
 data collected
by the DELPHI experiment during the period at high (LEP2) and low (LEP1) energy 
of the LEP operations.

\section{Results}
\label{par:results}
\begin{figure}[tbhp]
\centerline{
\epsfxsize=10.0cm\epsffile{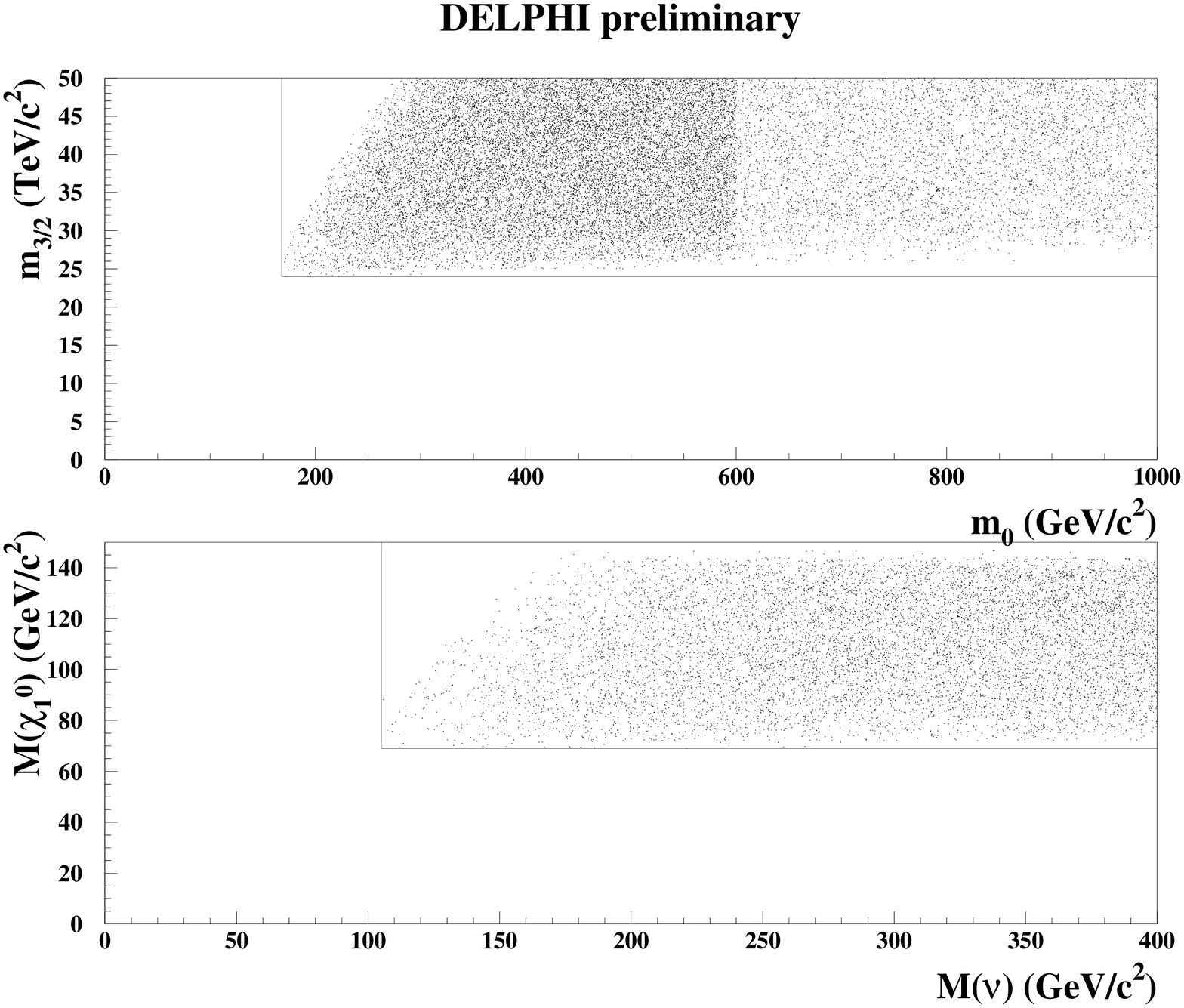}}
\caption[]{(Up) Physically allowed \mzero\ and \mhf\ parameters in AMSB  after having considered all the results of the searches. 
          (Down)Physically allowed \MXN{1}\ and \msnu\ masses in AMSB  after having considered all the results of the searches.}
\label{fig:isajet-mm}
\end{figure}

The searches results used in the present work, include  the  search  for  nearly  mass degenerate 
chargino  and  neutralino (always  present  in  AMSB), the search for Standard Model like or invisible Higgs boson, the search for stable staus, 
and the search for cascade  decays  resulting in the LSP (neutralino or sneutrino) and a low multiplicity final state containing neutrinos.
They   are fully described in the DELPHI AMSB~\cite{amsbex} paper and references therein.
In any of the searches, no excess of candidates was observed  with respect to the standard model predictions 
and limits on masses and AMSB theoretical parameters were set at 95 \% confidence level using {\tt ISAJET}~7.58~\cite{isajet} to compute the AMSB mass 
and cross-section spectra. Fig.~\ref{fig:isajet-mm} show, the remaining allowed points in the planes (\mzero,\mhf)(Up) and  (\msnu,\MXN{1})(Down) 
after having applied all the results of the searches  described in~\cite{amsbex}. 

One sees from Fig.~\ref{fig:isajet-mm}, that a gravitino lighter than 24~\TeVcc, a lightest neutralino lighter than approximately 69~\GeVcc\ and 
sneutrinos lighter than 105~\GeVcc\ should  not be allowed in AMSB. Moreover, a limit on the theoretical parameter, $m_0$ arising from the non-anomalous 
contribution is set to be above 168~\GeVcc\ at the electroweak scale independently of the breaking mechanism.
 
\section*{References}
\begin{thebibliography}{99}
\bibitem{branes}
 L. Randall, R. Sundrum, \Journal{\NPB}{557}{79}{1999}.
\bibitem{amsb}
 G.F. Giudice, M. Luty, H.Murayama, R.Rattazzi, JHEP {\bf 98} (1998) 12. 
\bibitem{amsbex}
 T. Alderweireld  \etal, DELPHI note 2002-06 CONF 547.
\bibitem{isajet}
 H. Baer, F.E. Paige, S.D. Protopopescu, X. Tata, ``Simulating Supersymmetry with ISAJET~7.0/ISASUSY~1.0'',
 Published in Argonne Accel.Phys.1993:0703-720.
\end{thebibliography}
\end{document}